# Nanosecond motions in proteins impose bounds on the timescale distributions of local dynamics

**Osman Burak Okan, Ali Rana Atilgan, Canan Atilgan**

Faculty of Engineering and Natural Sciences, Sabanci University, 34956 Istanbul, Turkey

*e-mail*: canan@sabanciuniv.edu
*telephone*: +90 (216) 483 9523
*telefax*: +90 (216) 483 9550

**ABSTRACT**

We elucidate the physics of the dynamical transition via 10-100ns molecular dynamics simulations at temperatures spanning 160-300K. By tracking the energy fluctuations, we show that the protein dynamical transition is marked by a cross-over from non-stationary to stationary processes that underlie the dynamics of protein motions. A two-time-scale function captures the non-exponential character of backbone structural relaxations. One is attributed to the collective segmental motions and the other to local relaxations. The former is well-defined by a single-exponential, nanosecond decay, operative at all temperatures. The latter is described by a set of processes that display a distribution of time-scales. Though their average remains on the picosecond time-scale, the distribution is markedly contracted at the onset of the transition. The collective motions are shown to impose bounds on time-scales spanned by local dynamical processes. The non-stationary character below the transition implicates the presence of a collection of sub-states whose interactions are restricted. At these temperatures, a wide distribution of local motion time-scales, extending beyond that of nanoseconds is observed. At physiological temperatures, local motions are confined to time-scales faster than nanoseconds. This relatively narrow window makes possible the appearance of multiple channels for the backbone dynamics to operate.

# Introduction

Protein function is made possible through fluctuations of varying amplitude and duration around the folded conformation of the molecule (1-3). In addition to its instrumental role in imposing the final folded state, the solvent forms a vicinal layer around the protein, mediating the functionality-related molecular fluctuations (4, 5). Therefore, a functional protein might well be perceived as an effective protein-solvent complex (6, 7). Proteins are characterized as soft materials (1, 8) and the temperature window for effective softness is bracketed below by the dynamical transition and above by the unfolding event (2, 9-11). The exact value of the dynamical transition temperature depends on environmental factors such as the solvent, the degree of solvation, in addition to the amino acid composition of the protein of interest.

The microscopic interpretation of the dynamical transition is subtler than the unfolding counterpart, the latter being apparent with the overall loss of the folded topology. To interpret the former, many attempts have been made, some concentrating on the solvent around the protein (4, 5, 12), others on the protein itself (9, 13). Furthermore, recent work has demonstrated that the dynamical transition occurs in the absence of protein tertiary or secondary structure, but is rather related to side chain interactions with the solvent (14). For a fully hydrated protein, the collisions with the surrounding vicinal layer of solvent molecules excite the protein by transferring energy which in turn is also dissipated by the viscous drag of the solvent (15-17). Recently, it has been suggested that solvent-slaved $\alpha$ relaxations, due to the strong coupling between surface residues and water, result in a large solvent reorganization energy, facilitating the functionality of the protein (18). The properties of the system below the dynamical transition are attributed to the existence of non-communicating local sub-states, which are postulated to interconvert at physiological temperatures. For instance, for parvalbumin and using X-ray crystallography, well-defined alternate conformations have been observed within the hydrophobic core at low temperatures, whose population-weighted average corresponds to the single conformation observed at room temperature in the same crystal form (19).

The temporal memory loss in the system depends on the dynamical processes that annihilate local changes. These may be recovered from the correlation function formalism (20), coupled with an appropriate mathematical model describing the underlying physics. Single exponential decays are the natural starting point for analyzing relaxation phenomena, since they are the phenomenological descriptors of simple dashpots, i.e., dampers which display frictional resistance to motion (15). However, the existence of non-exponential decays in molecular systems is now well recognized, supported by experimental and simulation data (21-25). The multi-exponential analysis, which extends the single exponential scope by adding more terms of the same form, is the natural next step (26). Due to the difficulty of assigning a physical meaning to the time scales obtained from a multi-exponential analysis, the stretched exponential function is adopted as an alternative (27, 28). In this approach, the relaxation profile due to many distributed time-scale processes, each contributing to the relaxation phenomena, are represented with a single characteristic time-scale and a dimensionless exponent that captures their distribution profile. Such behavior is observed when there are multiple channels for the process to decay (29).

In this study, we aim to answer two major questions: (i) What are the time and length scales of the processes that dynamically operate on the folded protein so as to render it functional above the dynamical transition; (ii) How do processes that operate on different time scales affect each other; in particular, do slower time-scale processes impose limitations on the fast, local motions, so as to provide coupled processes and/or multiple decay channels? To answer these questions, we elucidate the physics of protein dynamical transition via 10-100 ns long MD simulations on hen egg white lysozyme in water at a series of temperatures where the protein remains folded. By monitoring the energy fluctuations in the system, we show that the dynamical transition is marked by a cross-over from non-stationary to stationary trajectories. By further monitoring the backbone dynamics, we show that a slower, nanosecond, time scale motion in the protein imposes bounds on the distribution of the plethora of fast, local processes averaging on the picosecond range. Implications for controlling protein functionality are discussed.

## Methods

**Simulation Protocol.** The NAMD package is used to model the dynamics of the protein – water systems. The protein is soaked in a solvent box such that there is at least a 5 Å layer of solvent in each direction from any atom of the protein to the edge of the box. The simulated protein-water complex is comprised of 2271 TIP3P water molecules soaking a single hen egg white lysozyme [protein data bank (30) code 193L (31)] molecule, a 129 residue protein. Eight chlorine ions are added for charge neutrality. The system has a total of 10281 atoms, prepared using the VMD 1.8.6 program with solvate plug-in version 1.2 (32). The CharmM27 force field parameters are used (33). Electrostatic interactions are calculated via a partial mesh Ewald method (34). The cutoff distance for non-bonded van der Waals interactions was set to 12 Å with a switching function cutoff of 10 Å. Rattle algorithm was used to fix the bond lengths to their average values.

The system is first subjected to energy minimization until the gradient tolerance is less than $10^{-2}$ kcal/mol/Å with the conjugate gradients method. The resulting system with the orthorhombic periodic cell of starting dimensions $48 \times 45 \times 54$ Å is used as the initial structure in all MD simulations. All runs are made in the NPT ensemble at 1 atm with the NAMD software (35). Volumetric fluctuations are preset to be isotropic. Temperature control is carried out by Langevin dynamics with a dampening coefficient of 5/ps and pressure control is attained by a Langevin piston. Verlet algorithm is used for integrating the equations of motion with a time-step of 2 fs (36). 10 to 15 ns long MD trajectories are produced, spanning the temperature range 160 – 300 K with 10 K increments. In addition, 100 ns long trajectories are calculated for systems at 160, 180 and 300 K.

The energy trajectories are monitored to select the correct equilibration period for the MD runs. We find that at the end of 0.4 ns, all trajectories obtained at temperatures including and above 200 K display energy fluctuations occurring around a baseline average energy value. For temperatures spanning the 200-300 K range, all runs are therefore equilibrated for 400 ps and structures are recorded for 9.6 ns for subsequent data analysis. At lower temperatures, such convergence occurs in a much longer time period, and we have monitored each trajectory to use 9.6 ns of energy equilibrated data. This requires at least 5 ns time of equilibration period at 160-190 K. The energy trajectories of all the equilibrated systems are displayed in the supplementary material figure S1.

**Characterization of the Dynamics via $C_\alpha$ Positional Fluctuations.** Temporal dynamics of lysozyme is analyzed through positional autocorrelation functions of $C_\alpha$ carbons of each residue (figure 1). We and others find these to well-represent the ps – ns dynamics undergone by the folded protein (5, 37). As positional fluctuations occur through local deformations of the protein backbone, rigid body translations and molecular rotations are discarded to monitor the internal positional fluctuations only. Thus, the X-ray structure coordinates of lysozyme are treated as the reference and every recorded conformation from the MD trajectories is superimposed on it. The superimposition is based on the minimization of the quaternion norm pertaining to the differences in the atomic coordinates between the target (i.e. reference) structure and the MD recorded one (38). The $N \times 3$ coordinate matrix $\mathbf{R}(t)$ is thus generated at each time step, $t$. A mean structure is defined as the average over these coordinate matrices at each temperature, $\langle \mathbf{R}(t) \rangle$. One can then write the positional deviations for each residue $j$ as a function of time and temperature:

$$\Delta \mathbf{R}_j(t) = \mathbf{R}_j(t) - \langle \mathbf{R}_j(t) \rangle \tag{1}$$

The positional autocorrelation function of the protein at a given temperature is an average over that calculated for all residues $j$ as;

$$C(t) = \overline{C_j(t)} = \overline{\left[\frac{\langle \Delta \mathbf{R}_j(0) \cdot \Delta \mathbf{R}_j(t) \rangle}{\langle \Delta \mathbf{R}_j(0)^2 \rangle}\right]} \tag{2}$$

where the overbar represents an average over all residues, and the brackets are time averages.

In previous work, the initial decay (up to 50 ps) of the relaxation curves were found to be well-characterized by a stretched exponential form, which was originally proposed to represent the muti-exponential character of the relaxation phenomena (2, 9). In this approach, the relaxation profiles of $m$ distributed time-scale processes, each contributing to the relaxation phenomena with weight $a_i$ and individual relaxation time $\tau_i$, are approximated by,

$$C(t) = \sum_{i=1}^{m} a_i \exp\left(\frac{t}{\tau_i}\right) \approx \exp\left(\frac{t}{\tau}\right)^{\beta} \tag{3}$$

where $\tau$ is a characteristic time scale and $\beta$ is a dimensionless exponent. The physical basis of $\beta$ has not been unambiguously established. Yet, it is known to carry the information of superposed single exponential decays for the range $0 \leq \beta \leq 1$ following from its Laplace transform.(39)

A time-decay that may be interpreted via a stretched exponential function involves a dashpot with a time-dependent damping coefficient. The relaxation times obtained from fits to a stretched exponential model are scaled with the $\beta$ exponent; for $\beta < 1$ they effectively extend into time scales slower than $\tau$. The initial infinite decay rate of a stretched exponential function follows from its power-law descriptor (40), and is physically ascribed to a purely elastic response preceding a viscous decay, accounting for the elastic contribution.

We find that it is not possible to model the total decay of the $C_\alpha$ relaxations using the stretched exponent model as the single term. While that model captures the initial decay profile faithfully, it misrepresents the long-time tail. The latter is well-captured by a single exponential decay model. Here, we propose a new functional form for non-exponential $C_\alpha$ relaxations by superposing a single exponential term on the stretched exponential model;

$$C(t) = a \exp\left[-(t/\tau_f)^{\beta}\right] + (1-a) \exp[-(t/\tau_s)] = C_f(t) + C_s(t) \tag{4}$$

so as to capture the overall relaxation profile. This treatment will provide additional information to our previous studies where the focus was on the initial relaxations on the picosecond time scales (2, 5, 9). In equation 4, $C_f$ and $C_s$ are the contributions to the overall correlations from the fast and slow processes, respectively; $\tau_f$ and $\tau_s$ are the corresponding characteristic times. The positive front factor, $0 \leq a \leq 1$, describes the relative contributions from processes at these separated time scales. This composite model reflects the properties of both the short and long-time behavior of the relaxations as well as the crossover, as will be shown in the **Results** section.

In each calculation, we first proceed to determine the relaxation curves from the MD trajectories, as defined in equation 2. The 9.6 ns long trajectories are used in the calculations yielding smooth curves, which are found to have completely relaxed to zero within 2 ns at all temperatures. We then find the best curve-fit using the four parameter model in equation 4. The fitting procedure used is based on an exhaustive search in the space of permitted values of $a$, $\beta$, $\tau_f$ and $\tau_s$. Initially, $a$ and $\beta$ are incremented in units of 0.1 in the interval [0,1], $\tau_f$ is incremented in units of 2 ps in the interval [2,50], and $\tau_s$ is incremented in units of 50 ps in the interval [50,1500]. The optimal solution is identified as the global least squares error minimizer between the fitted and calculated $C(t)$ curves. At different temperatures, the number of points used from the calculated $C(t)$ curves in the fitting procedure is actively adjusted to be in the range 1.0 – 1.6 ns. This prevents excessive weighting of long time tails at the expense of a poorly captured crossover region. Once different solutions are reached in this way, the final solution is verified to be the global least squares error minimizer over the 2 ns, fully relaxed $C(t)$ curve. For this purpose, different regions of optimality are bracketed for all four parameters, and they are further subjected to finer grained exhaustive search in increments of 0.01 for $\beta$, 1 ps for $\tau_f$ and 10 ps for $\tau_s$.

In closing, we note that we have also tried an extended five-parameter model, where the second term in equation 4 includes an additional stretched exponent. We found that the exponent due to the slow processes always converges to 1 at all temperatures, reducing to the model in equation 4. Thus, we find that the $C_\alpha$ positional fluctuations decay with complex dynamics at short (ps) times, whereas their longer (ns) time tails are well-represented by a single exponential decay.

## Results

**Stationary Behavior and the Dynamical Transition.** For a stationary process, both the averages (e.g., the energy), and the fluctuations of a given physical observable (e.g., the heat capacity) should be independent of the window length of the trajectory used. However, especially at low temperatures, such stationarity may not have been achieved; thus, the equilibration observed in figure S1 does not directly imply stationarity. We therefore calculate the average energies over different window lengths, $w$, and find that these averages are predominantly $w$ dependent below 200 K, and are independent of $w$ above 200 K. Sample trajectories at 180 and 300 K are shown in figure S2, where the averages over window lengths of $w$ = 200, 400, 800, 1600 ps are marked.

We further investigate the fluctuations in energy through the heat capacity. Previously, we have monitored the heat capacity of several protein-solvent systems to determine the dynamical transition (5, 9). In those studies, trajectories were partitioned into 200 ps long chunks, and the heat capacity was calculated for each chunk from the energy fluctuations from, $(\langle E^2 \rangle - \langle E \rangle^2)/k_B T^2$, where $E$ is the instantaneous energy at temperature $T$ and $k_B$ is the Boltzmann constant. The heat capacity at a given temperature was then reported as an average over those obtained in different chunks. To characterize the presence/absence of stationary behavior in the system at different temperatures, we display in figure 2, the heat capacity calculated from trajectories of length 9600 ps, by partitioning them into lengths of $w$ = 200 (48 sets), $w$ = 400 (24 sets), $w$ = 800 ps (12 sets), and $w$ = 1600 ps (6 sets). We find that at and above 200 K, the value of the heat capacity is the same for all trajectory lengths within the error bounds (the standard error on the mean of the data, $\varepsilon$, averaged over the $n$ = 48 sets obtained with $w$ = 200 ps is shown by the shaded area, with $\varepsilon = \sigma/\sqrt{(n-1)}$, where $\sigma$ is the standard deviation and $n$ is the sample size).

Conversely, at lower temperatures, it depends strongly on the window length of the trajectory used in the calculation, with larger fluctuations for larger $w$.

We postulate that at these lower temperatures, the system may only achieve piecewise stationarity. For the energy landscape, this would imply a system settling at local minima corresponding to metastable states, with occasional jumps between those states; these local minima, however, decorate a larger energy well that is the attractor for the whole system (41). This type of non-stationary behavior may be associated with glassiness for which the concept of metastable states is central (22).

Thus, we find that it is possible to define the dynamical transition as the point where the non-stationarity in the data disappears. This transition may be uniquely distinguished by probing the temperature below which the heat capacities obtained from various time windows do not display a convergent behavior. Note that such a measure of the dynamical transition is suggested for the first time in the literature. We will show that the transition from the non-stationarity to the stationarity in the time evolution of states has implications for the time scales governing the backbone dynamics of the protein. We next set out to characterize the origin of the various contributions to the protein-water dynamics below and above this transition temperature.

**Characterization of the Backbone Dynamics.** In the folded protein, although the backbone goes through incessant fluctuations, there are no conformational changes, as would be characterized by jumps in the backbone torsional angles ($\varphi, \psi$). This is contrary to the side chain torsional angles $\chi$, where occasional jumps between the conformational states are recorded, in particular in the surface groups and at temperatures above the dynamical transition (5, 9). The former is prohibited by the free-energetic cost of a required counter-rotation in a nearby torsion in polymeric chains to prevent large displacements of chain ends [see (42) and references cited therein]. Thus, the backbone dynamics is ideal for reflecting the contributions of the ps-ns time scales on the overall dynamics of the system due to the well-defined "average" for this observable.

By applying equation 2, we first study the average $C_\alpha$ relaxations over different time windows, $w$; *i.e.*, the time averaging is done for different lengths of trajectories. We make use of the $W = 100$ ns long trajectories generated at 180 K and 300 K, and partition them into $m = \text{int}(W/w)$ sub-trajectories each of which we consider to be independent so far as the backbone relaxations are concerned. We find that the relaxation depends on the time-window of observations as displayed in figure 3 for the system at 300 K. At short time windows of observation, up to the order of a few nanoseconds, the relaxation curves shift to incorporate higher time scales. In the observation window of $w = 7.2 – 24$ ns, however, the relaxation profiles are nearly the same. This observation is interpreted as follows: At certain observation windows there are some partially relaxed processes contributing to the decays. Thus, by elongating the observation time, $w$, a larger portion of these are included in the decay profiles. For certain $w$, however, all contributing processes have essentially relaxed, forming a plateau. Above $w = 24$ ns, the relaxation curves display shifts to higher relaxation times, indicating that longer time scale processes begin to influence the dynamics in these longer windows. We note that the same results are reproduced at 180 K, indicating that this behavior is independent of the state of the system. To better treat the physics of fully relaxed processes, we choose $w = 9.6$ ns at all temperatures studied. We find that equation 4 is an excellent model fit to all the curves. It represents the overall relaxation behavior, including the initial relaxation, the long-time tail and the crossover between the two regimes. A sample relaxation curve of the backbone fluctuations is provided in the supplementary material figure S3. We display *C*(*t*) at 180 K (below the dynamical transition) computed via equation 2, and the model fit of equation 4; the short time region is magnified in the inset. The short time decay depicts a sharp elbow below the dynamical transition, which smoothes out at higher temperatures.

The results of the fits to equation 4 are displayed for the exponent *β* in figure 4a. MD results imply a biphasic character for the exponent, best described by a Boltzmann sigmoidal function as in our previous work (9). Levels are at ca. *β* = 0.2 below the dynamical transition and ca. *β* = 0.4 above it; the transition temperature is predicted at 192 ± 2 K. Inasmuch as *β* describes the initial decay characteristics of the relaxation, these results are the same as those obtained in our previous work for the proteins BPTI (9) and endogluconase in water and glycerol (5), whereby only the initial decay characteristics were studied.

We further display the characteristic times of the fast and slow processes in figure 4b. We find that $\tau_s$ is independent of temperature, with an average value of 0.8 ± 0.1 ns. On the other hand, $\tau_f$ shows an increase from *ca.* 2 ps below the dynamical transition to 11 ± 1 ps. We have previously explained this slowing-down in the local relaxations by a model that incorporates the onset of the coupling between the local internal motions of the protein and the vicinal solvent layer with temperature (5). We have shown that such behavior is due to interplay between the decreased stiffness and the modified effective friction coefficient.

In sum, we characterize the backbone dynamics of the folded protein in the temperature range of interest and up to time scales of ca. 20 ns. It accommodates the superposition of fast local relaxations with a wide distribution of time scales ($\tau_f \approx 2 - 10$ ps, $\beta \approx 0.2 - 0.4$) and a more collective, single time scale motion ($\tau_s \approx 0.8$ ns). The relative contribution of the slow and fast time scales to the overall relaxation remains the same at all temperatures (*i.e.*, $a \approx 0.74 \pm 0.02$).

In the previous subsection, we have shown that before the onset of the dynamical transition, the trajectories are non-stationary. Communication between the different local sub-states of the protein occur in a discrete manner, displaying a widely distributed range of relaxation times, characterized by $\beta \approx 0.2$ and a picosecond time scale of $\tau_f$. With the onset of the transition, continuous communications between the different local sub-states emerge, accompanied by the increase in *β* to *ca.* 0.4. The latter indicates a change in the distribution of relaxation times of the fast processes. The more collective motions are ever-present on the nanosecond time scale ($\tau_s$). The question arises as to how, if at all, these slow time-scale motions influence the fast internal processes, which we elaborate upon in the next subsection.

**Slow collective motions in proteins impose bounds on the distribution of time scales of fast processes.** The interpretation of the long observed trend for the increase from $\beta \approx 0.2$ to 0.4 for the distribution of the local time scale motions with the onset of the dynamical transition (2, 5, 9) may now be rationalized. The overall effect of *β* is made transparent when the characteristic relaxation time of the total decay, *τ**, is studied. We define the integrated area *τ** of the autocorrelation function as the average time constant of the relaxation (40). Using the model in equation 4 this yields:

$$\tau^* = \int_0^\infty C(t)dt = a\frac{\tau_f}{\beta}\Gamma\left(\frac{1}{\beta}\right) + (1-a)\tau_s \qquad (5)$$

where the solution involves the gamma function, $\Gamma(1/\beta)$. Due to the mild dependence of $\tau_f$ as well as the independence of *a* and $\tau_s$ on temperature, the change in the overall relaxation behavior is governed by the stretched exponent as displayed in figure 4. Thus, $\tau^*$ substantially decreases with onset of the dynamical transition, from ca. 1 ns to 200 ps (figure 5).

$\beta$ values in the range of 0.2 at low temperatures render much faster time scale motions compatible with the slow one via an effective stretching of time. From another viewpoint, the distribution of the time scales of the fast processes at these temperatures are such that their effect is not felt by the overall system, so that the slow process time scale emerges as the controlling factor, decoupled from the local motions, so that $\tau^* \approx \tau_s$. At temperatures where proteins are functional, on the other hand, there is substantial coupling between the collective motions, and the more localized picosecond time scale degrees of freedom, so that $\tau_f < \tau^* < \tau_s$.

We find that the coexistence of the slow time scale belonging to the collective motions with a couple of orders of magnitude separation imposes both an upper and a lower bound on these values, along with an understanding of how the protein and its environment operate. These are discussed below:

The upper bound on $\beta$ is observed at physiological temperatures. Since there are a large number of fast exponential decaying processes contributing to the local dynamics in the protein environment, we assume a continuous spectrum of relaxation processes for the stretched exponential component. The distribution function of the stretched exponential, $\rho(\tau,\beta)$, may be expressed as a convergent infinite summation (43):

$$\exp(-t/\tau_f)^\beta = \int_0^\infty \exp(-t/\tau)\,\rho(\tau,\beta)d\tau$$
$$\rho(\tau,\beta) = -\frac{\tau_f}{\pi\tau^2}\sum_0^\infty \frac{(-1)^k}{k!}\sin(\pi\beta k)\Gamma(\beta k+1)\left(\frac{\tau}{\tau_f}\right)^{\beta k+1} \qquad (6)$$

Here $\tau$ covers the processes distributed with all possible characteristic times. For $\beta = 1$ the sum reproduces the Dirac delta function centered on $\tau_f$. The probability of finding processes with time scales extending into very slow regimes increases as $\beta \to 0$. In figure 6, we display for ($\beta = 0.2$, $\tau_f = 2$ ps) and ($\beta = 0.4$, $\tau_f = 11$ ps), the distribution functions $G(\tau,\beta) = \tau\,\rho(\tau,\beta)$ which is the form customarily used to present these distribution functions (43). The area under the curves are normalized, so that $\int_{-\infty}^\infty G(\tau,\beta)d\ln\tau = 1$.

We find that the spectrum of fast relaxation times at $\beta = 0.4$ effectively vanishes at $\tau = 1$ ns $\approx \tau_s$. On the other hand, for $\beta = 0.2$ the distribution has appreciable contributions from a much wider range, extending into the 10 ns time scales. Therefore, $\beta = 0.4$ forms an upper bound for delimiting the distribution of time scales at nanoseconds.

Note that equation 6 assumes a continuous distribution over an infinite number of contributing processes. In practice, a low $\beta$ value is attainable even with two processes, with the condition that their time scales are well-separated (9). In proteins, we expect to find a large, but finite number of fast time scale motions, and the above discussion merely points to the fact that these are widely distributed over many orders of magnitude below the dynamical transition. On the other hand, they are confined to a smaller region at higher temperatures, enabling their efficient coupling with the collective motions. In other words, instead of there being very slow or very fast isolated relaxation events, they are confined to a narrower region where communication between different processes is made possible.

To rationalize the lower bound of $\beta \approx 0.2$ observed at low temperatures, we rely on a model of hierarchically constrained dynamics, accounting for the slowing down in hierarchically ordered materials such as glasses and proteins (44-46). In this model, the local energy landscape is viewed as consisting of microstates. The system has to overcome a succession of energy barriers of increasing heights, belonging to the $n$ sub-states decorating each microstate. From the ongoing discussion related to figure 2, we conjecture that below the dynamical transition, the non-communicating local sub-states lead to such hierarchy with a wide distribution of time scales. The assumption of non-communication is based on the observed non-stationary behavior of the trajectories.

The characteristic time of each contributing process $i$ must overcome the barrier $\Delta E_i$ by the Arrhenius law $\tau_i \sim \exp(\Delta E_i / kT)$, where $k$ is the Boltzmann factor. Since the dynamical model relies on the hierarchical sampling of states, the total energy spent in a microstate with $n$ sub-states is $\Delta E_n = \Sigma_i \Delta E_i$. The total time spent in a given microstate, $T_n$, is then a series of multiplications of the time scales of the contributing sub-states: $T_n \sim \Pi_i \tau_i$. An observable relaxing macroscopically is given by $\langle \exp(t/T_n) \rangle$ where the average is carried out over the distribution of $T_n$, due to the presence of the many microstates, each decorated with $n$ sub-states. Using extreme deviation theory, it is shown (44, 45) that this average may be expressed as the stretched exponential, $\exp(t/\tau)^\beta$. Furthermore, the integral in equation 5 becomes:

$$\tau^* = a \frac{\tau_f}{\beta} \Gamma\left(\frac{1}{\beta}\right) + (1-a)\tau_s = a\langle T_n \rangle + (1-a)\tau_s = \tau_s + a[\langle T_n \rangle - \tau_s]$$

Thus, one can write the following:

$$\tau^* \sim \exp\frac{\Delta E_s}{kT} + a\left[\langle \exp\frac{\Delta E_n}{kT}\rangle - \exp\frac{\Delta E_s}{kT}\right]$$

where, again, the average is carried out over the distribution of $T_n$. The energy associated with the convex hull of the $n$ sub-states is maintained by the prevalent nanosecond time scale collective motions of the protein. The situation where the average energy spent in the microstates is compensated by that of the slower relaxations of the hull forms a limiting case. This limiting case sets the last term in square brackets to zero. Equating the two contributing terms, with $\tau_f = 2$ ps, and $\tau_s = 0.8 \pm 0.1$ ns, gives the limiting value of $\beta$:

$$\frac{\tau_f}{\beta}\Gamma\left(\frac{1}{\beta}\right) \approx \tau_s \quad \rightarrow \quad \beta \cong 0.18 \pm 0.01$$

Smaller values of $\beta$ give explosively slower overall relaxation times ($\tau^*$), which would let the fast processes take control over the whole dynamics; values larger than 0.2 yield small contributions to the sum in equation 5 so that $\tau^*$ is driven towards the sub-nanosecond time scale. Thus, the protein-water system below the dynamical transition may be assumed to have converged to this limiting behavior. Note that, above the dynamical transition, the hierarchical model is no longer applicable, and the observed multi-exponential decay may, for example, be explained by sub-diffusive behavior (47).

## Concluding Remarks

In this study, we have studied backbone dynamics of folded proteins spanning time scales on five orders of magnitude and in a wide temperature window, to gain a general understanding of the dynamics. We show that the dynamical transition in the protein results in part from the various time scales coming into concert, allowing large structural motions to occur through the cooperativity of the various accompanying local motions.

The view that proteins exhibit a hierarchical arrangement of conformational sub-states due to structural constraints is well established (48, 49). The presence of tiers along the conformational energy surface of proteins has also been exploited experimentally to study various dynamic properties. For instance, Samuni *et al.* have studied kinetic allosteric intermediates and equilibrium populations of hemoglobin (50) as well as the sequential onset of different tiers of dynamics in myoglobin (51). Their results were obtained under the assumption that in an environment that leads to the slowing down of time scales (*e.g.*, by the effect of viscosity, temperature, mutants), there will be a separation of time scales between the different processes contributing to the protein dynamics; those that are slaved the most will influence the dynamics with the largest delay.

Similarly in this work, as the system is cooled below the dynamical transition temperature, it is possible to resolve various sub-states that contribute to the dynamics, captured in the non-stationary nature of the trajectories. At these temperatures, where functionality has not been onset, the localized "fast" processes display heterogeneous dynamics with a characteristic time scale on the order of $2 - 10$ ps. Their distribution covers several orders of magnitude (figure 6) and even accommodates very slow relaxations with time-scales beyond that of nanoseconds with low probabilities.

We further find that below the dynamical transition temperature, the partitioning and the miscommunication of the local motions manifest themselves in the non-stationary behavior of the total energy trajectories. Communication between the sub-states is established above the transition temperature with the redistribution of the time scales of these local motions, as evidenced by the increase in the stretched exponent that characterizes these distributions (figure 4), and as detected in the heat capacity calculations (figure 2). A slower time scale on the order of 1 ns is present at all temperatures. We conjecture that this time scale belongs to segmental motions displaying homogenous dynamics.

Certain motions that only appear with the proper coupling of the local protein motions and the vicinal layer of solvent, such as side-chain motions (19, 52, 53), belong to the group of fast processes and take part in the redistribution of time scales (5). The ever-present slower dynamics acts as an agent that bridges the time scales. We argue that the coexistence of the slow time scale belonging to the collective motions imposes both an upper and a lower bound on these values, along with an understanding of how the protein and its environment operate. Thus, in environments that lead to the slowing down of time scales, certain motions may not just be delayed but may be completely hindered.

**Acknowledgements.** This work was partially supported by the Scientific and Technological Research Council of Turkey Project No. 106T522.

# Figure Captions

**Figure 1.** Sketch of the fluctuation vectors for $C_\alpha$ atoms. The protein conformations obtained from a chosen duration of the simulations are superimposed to obtain the average structure (depicted by the purple protein backbone); the instantaneous locations of the $C_\alpha$ atoms from 20 snapshots are shown as colored spheres. The fluctuation vector of a given residue at time *t* is the vectorial difference in the current position of the $C_\alpha$ atom and its location in the average structure. Three vectors for a chosen atom *j* are magnified on the lower left.

**Figure 2.** Heat capacity calculated from the energy trajectories using $(\langle E^2 \rangle - \langle E \rangle^2)/k_B T^2$. The different curves are obtained by partitioning the 9.6 ns trajectories into chunks of length 200, 400 800, and 1600 ps (48, 24, 12 and 6 sets, respectively.) The error bounds on the 200 ps window size calculations are shown with the gray shaded area. At physiological temperatures that emerge with the onset of the dynamical transition at 200 K, the length of the trajectory used does not affect the heat capacity values, within error bounds. Below 200 K, calculations depend on the length of data used, due to the non-stationary nature of the trajectories.

**Figure 3.** Relaxation curves, *C(t)*, at 300 K obtained for time windows of various length *w* = 1.2 ns to 36 ns. We observe that the relaxation curves shift in regions below *ca.* *w* = 7.2 ns and above *ca.* *w* = 24 ns. In the interim values of *w*, the general behavior of the relaxation profile remains the same. We therefore choose *w* = 9.6 ns (gray curve) as the observation window in all the relaxation curve calculations. Note that the same results are reproduced at 180 K. The same data are plotted on the log-log scale in the inset.

**Figure 4.** **(a)** Variation of *β* exponent with temperature, obtained from the fit of equation 4 to the calculated *C(t)* curves. A Boltzmann sigmoidal function fit to the data predicts the dynamical transition at 192 ± 2 K, and *β* levels at 0.19±0.01 below the dynamical transition, 0.37±0.01 above it. **(b)** The characteristic decay times for the fast and slow processes, $\tau_f$ and $\tau_s$ respectively, obtained from equation 4. The front-factor, *a*, that dictates the relative contributions of the two time scales to the overall relaxation phenomenon is temperature independent with $a \approx 0.74 \pm 0.02$.

**Figure 5.** The characteristic time scale, $\tau^*$, obtained from the combination of the fast and slow processes (equation 5). The line is an exponential fit to the data and is provided only as a guide to the eye.

**Figure 6.** Distribution function $G(\tau, \beta) = \tau \rho(\tau, \beta)$ for the fast time scale contribution to the relaxation with the characteristic time scale of $\tau_f$=10 ps. Curves for *β* = 0.2 and 0.4 are displayed. We find that the spectrum of relaxation times at *β* = 0.4 effectively vanishes at $\tau/\tau_f \cong 100$. On the other hand, for *β* = 0.2 the distribution has appreciable contributions from a much wide range, scanning the femtosecond – microsecond time scales. The area under the curves are normalized, so that $\int_{-\infty}^{\infty} G(\tau, \beta) d \ln \tau = 1$.

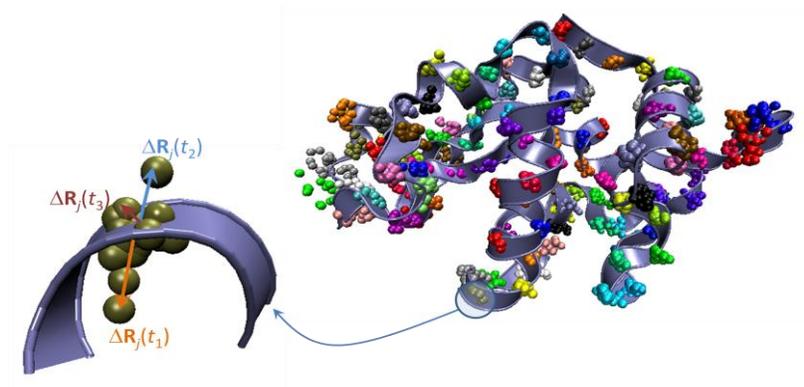

**Figure 1**

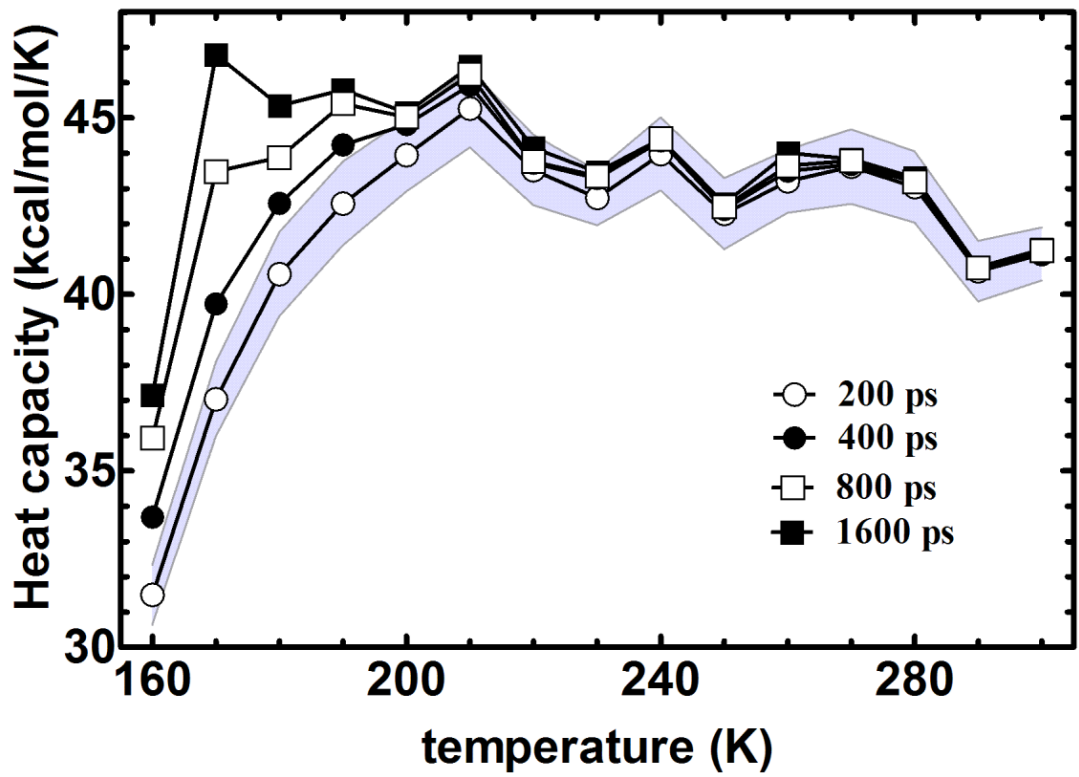

**Figure 2**

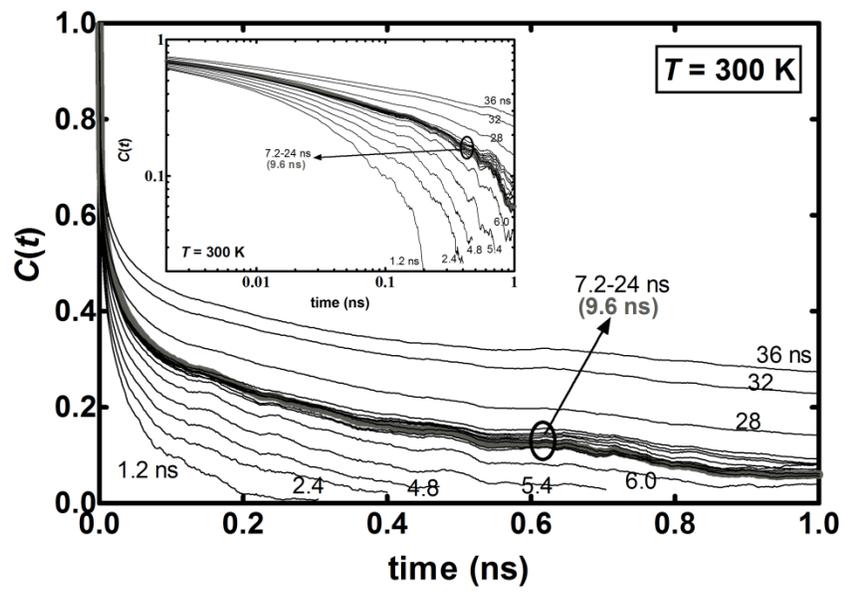

**Figure 3**

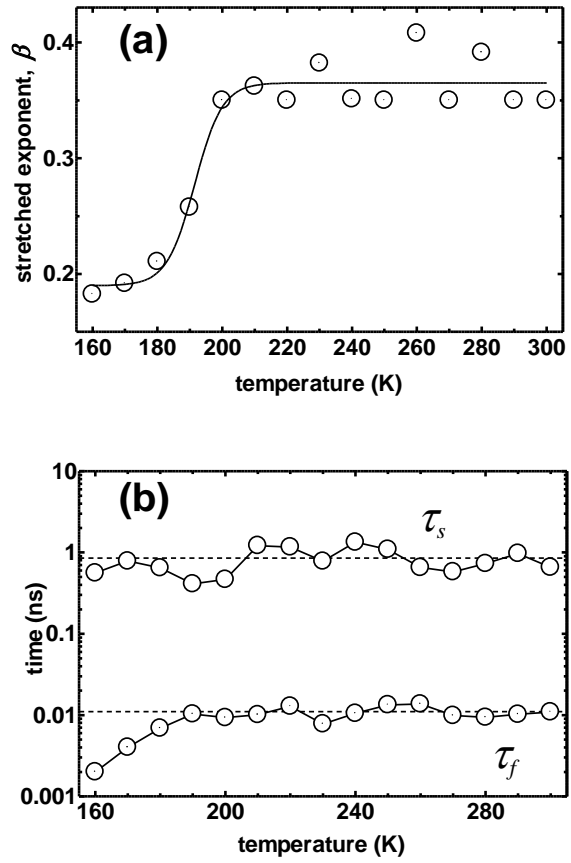

**Figure 4**

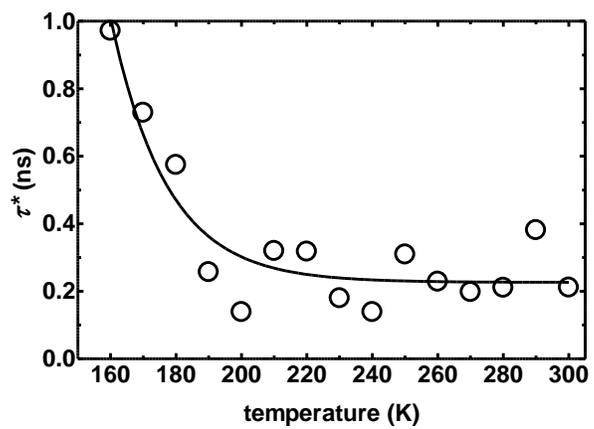

**Figure 5**

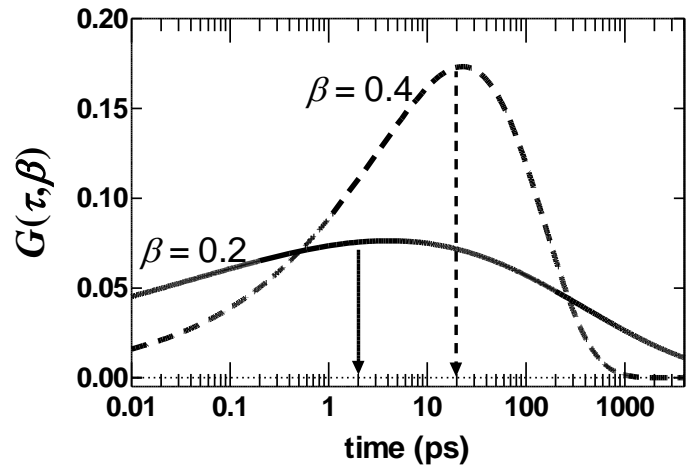

**Figure 6**

## Supplementary Material

**Figure S1.** 9.6 ns long energy trajectories for the equilibrated protein – water systems at all temperatures studied (listed on the right-hand-side). Equilibration periods at each temperature are not shown. These are at least 5 ns including and below 200 K. Above this temperature, 0.4 ns equilibration period is sufficient. The systems have reached equilibrium, and fluctuate around a baseline (average) value; the slope of each baseline is zero (< 0.01 kcal/mol/ps).

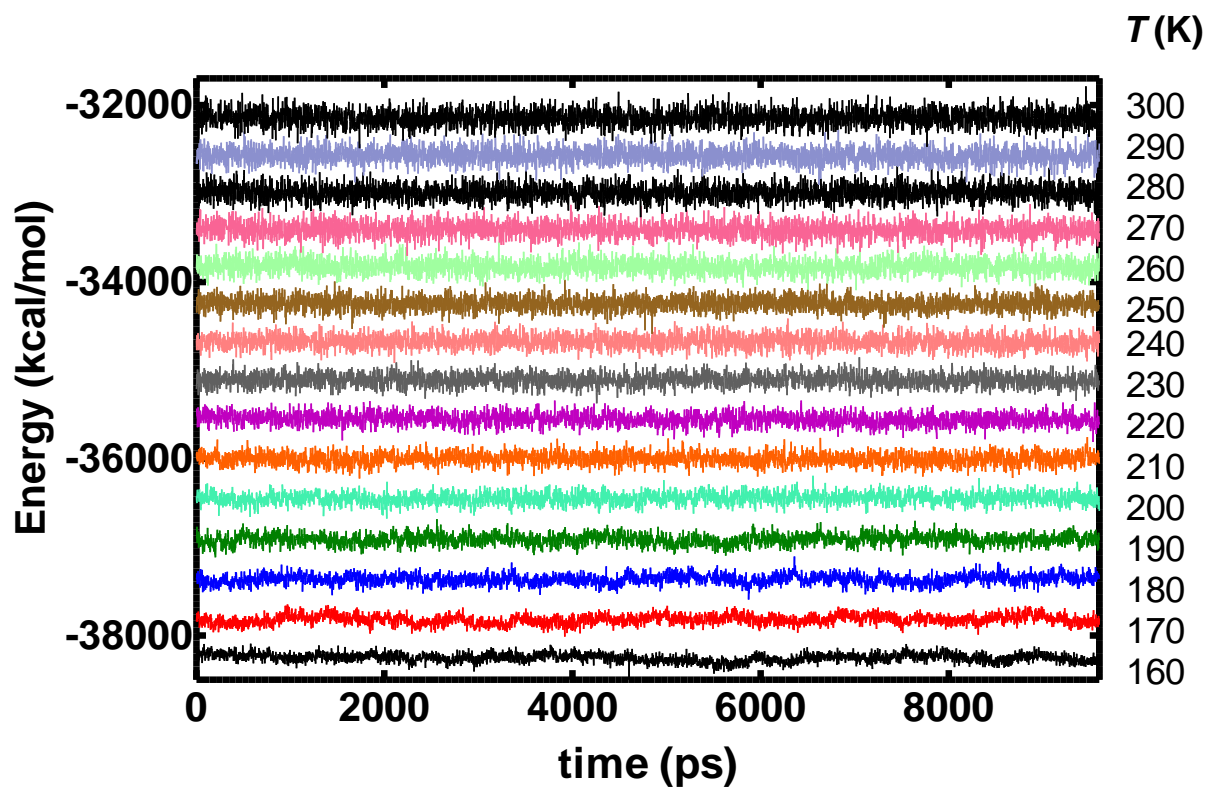

**Figure S2.** 8 ns pieces of sample, equilibrated energy trajectories, at 180 K (top figure) and 300 K (bottom figure). Marked on the figure are the average energy values for trajectory chunks of 200 ps (green), 400 ps (red), 800 ps (orange), and 1600 ps (blue), as well as the average over the whole trajectory (purple). At 180 K, significant deviations from the overall average are observed for all lengths of trajectory pieces, pointing to the non-stationary nature of the trajectory at this temperature. At 300 K, these deviations are minimal, showing that the energy trajectories are stationary.

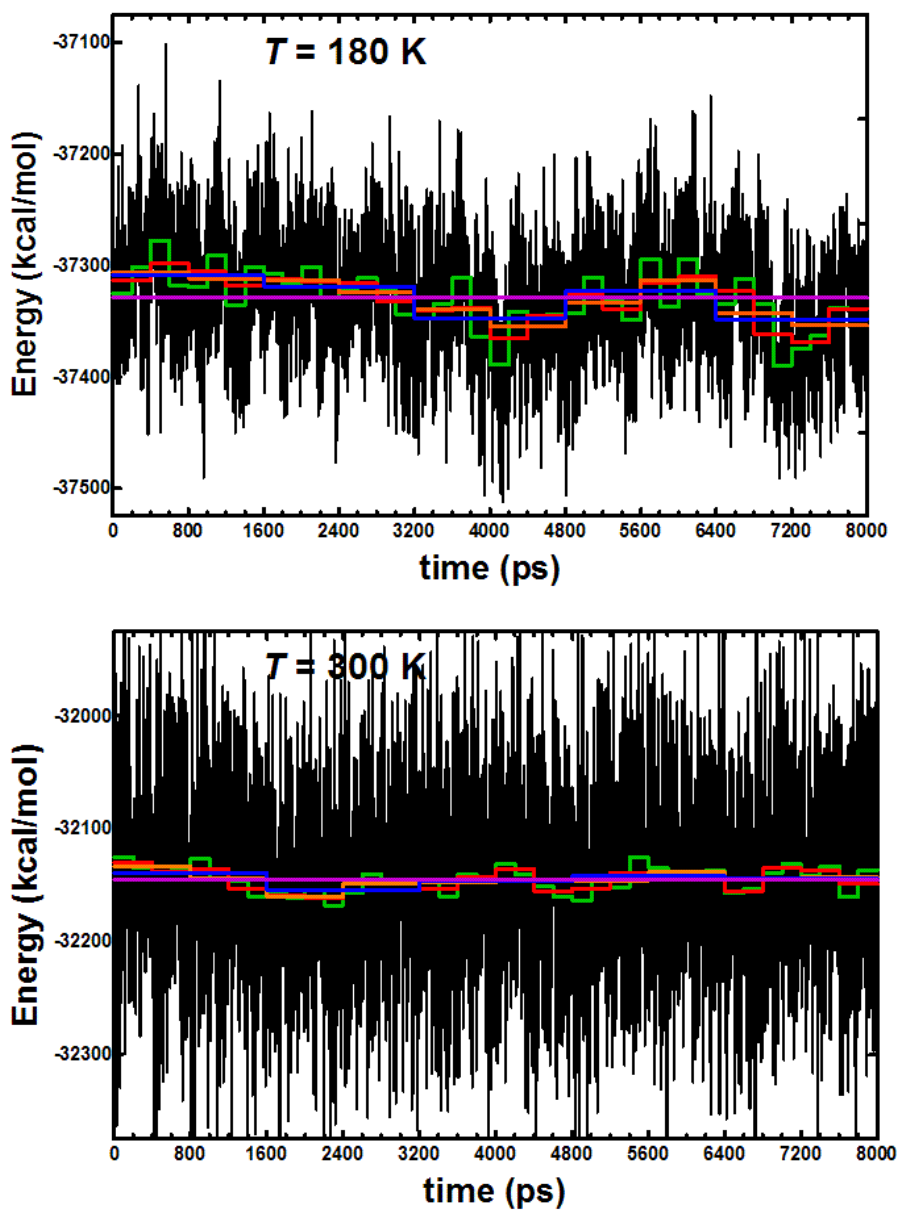

**Figure S3.** Sample relaxation at 180 K (black curve) and the fitted function using the model in equation 4 (gray dashed line). Full relaxation of the $C_\alpha$ fluctuations is observed within 2 ns. In the **inset**, the initial 100 ps part is magnified for better assessment of the goodness of fit at shorter timescales (upper curve). Also shown in the **inset** is the $C(t)$ curve obtained from an average over the 48 trajectory chunks of length 200 ps where the mean structure of equation 1 is obtained for each of the sub-trajectories (dashed line).

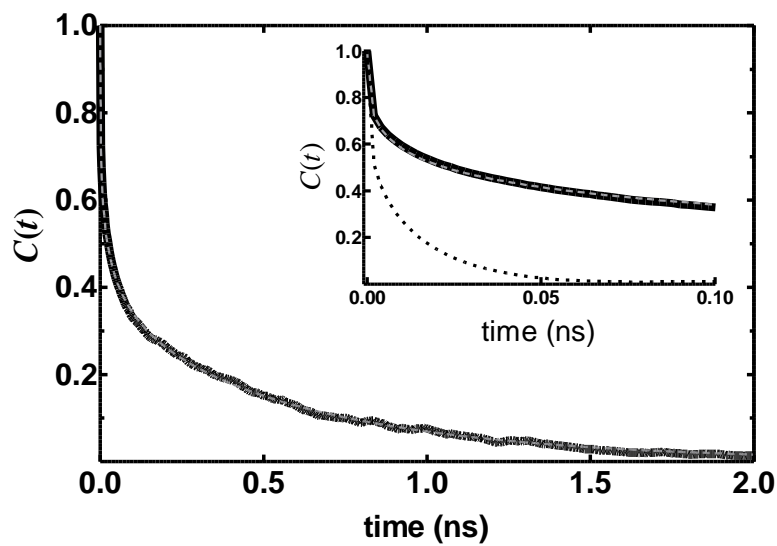